\documentclass[aps,pra,showpacs,groupedaddress,twocolumn]{revtex4}

\usepackage{graphicx}
\usepackage{dcolumn}
\usepackage{bm}
\usepackage[usenames,dvipsnames]{color}
\usepackage{amsmath}
\usepackage[hypertex]{hyperref}
\usepackage{amsfonts}
\usepackage{amsthm}
\usepackage{amssymb}
\usepackage{mathrsfs}
\usepackage{subfigure}
\usepackage{ulem}

\begin{document}

\title{Performance of various correlation measures in quantum renormalization-group method: A case
study of quantum phase transition}

\author{Yao Yao}

\author{Hong-Wei Li}

\author{Chun-Mei Zhang}

\author{Zhen-Qiang Yin}
\email{yinzheqi@mail.ustc.edu.cn}

\author{Wei Chen}

\author{Guang-Can Guo}

\author{Zheng-Fu Han}
\email{zfhan@ustc.edu.cn}
\affiliation{Key Laboratory of Quantum Information,University of Science and
Technology of China,Hefei 230026,China}

\date{\today}

\begin{abstract}
We have investigated quantum phase transition employing the quantum renormalization group (QRG) method while in most previous literature
barely entanglement (concurrence) has been demonstrated. However, it is now well known that entanglement is not the only
signature of quantum correlations and a variety of computable measures have been developed to characterize quantum correlations
in the composite systems. As an illustration, two cases are elaborated: one dimensional anisotropic (i) XXZ model and (ii) XY model,
with various measures of quantum correlations, including quantum discord (QD), geometric discord (GD), measure-induced
disturbance (MID), measure-induced nonlocality (MIN) and violation of Bell inequalities (eg. CHSH inequality). We have proved that all these
correlation measures can effectively detect the quantum critical points associated with quantum phase transitions (QPT) after several iterations
of the renormalization in both cases. Nonetheless, it is shown that some of their dynamical behaviors are not totally similar with entanglement
and even when concurrence vanishes there still exists some kind of quantum correlations which is not captured by entanglement. Intriguingly,
CHSH inequality can never be violated in the whole iteration procedure, which indicates block-block entanglement can not revealed by the CHSH
inequality. Moreover, the nonanalytic and scaling behaviors of Bell violation have also been discussed in detail. As a byproduct,
we verify that measure-induced disturbance is exactly equal to the quantum discord measured by $\sigma_z$ for general X-structured states.

\end{abstract}


\maketitle
\section{INTRODUCTION}
The origin of research into quantum correlation, or more precisely, quantum entanglement can date back to the EPR paper \cite{EPR} in 1935,
and nowadays it is no doubt that quantum entanglement is one of the most significant concepts in quantum information processing \cite{Nielsen}.
It has already been recognized as the fundamental feature of quantum mechanics and utilized as a crucial resource for communication and computation.
However, entanglement should not be viewed as the unique measure of quantum correlations since there exist other types of nonclassical correlations
which are not captured by entanglement. Recently many authors have proposed a variety of computable measures to characterize quantum correlations
in the composite states: quantum discord (QD) \cite{Ollivier2001,Henderson2001}, geometric discord (GD) \cite{Dakic2010,Luo2010a}, measure-induced disturbance
(MID) \cite{Luo2008a}, measure-induced nonlocality (MIN) \cite{Luo2011}, ameliorated MID \cite{Wu2009} and so on.
Within such a quantum-classical framework, a great deal of concern has been raised by quantum discord and discord-like correlation measures
in the past few years (for review, see \cite{Modi2011} and references therein).

In particular, as an important application in quantum phase transition (QPT) \cite{Sachdev}, entanglement can be exploited to determine the critical points
(CP) for spin chains at zero temperature \cite{Osterloh2002,Osborne2002,Vidal2003,Wu2004,Amico2008}. Meanwhile, since quantum discord (and other discord-like
measures) is introduced as an information-theoretical tool to qualify and quantify quantum correlations, it is natural for us to clarify the role played by
quantum discord in QPT. Several studies concerning such a relationship have already appeared in Ref. \cite{Dillenschneider2008,Sarandy2009,Werlang2010}.
These recent observations demonstrate that the CP information provided by QD is just in agreement with that of entanglement, and even at finite temperature
QD still works fine \cite{Werlang2010}.

Instead of resorting to two-point spin-spin correlation functions, which is usually done in most previous literature, quantum renormalization group
(QRG) method \cite{Wilson1975,Pfeuty1982} is introduced to investigate the quantum information properties of critical systems. Invoking such a method,
surveys regarding Ising and Heisenberg models have been carried out by several works \cite{Kargarian2007,Kargarian2008,Jafari2008,Kargarian2009,Ma2011a,Ma2011b}
and it has been shown that implementation of QRG method is valuable in detecting the nonanalytic behavior of entanglement (concurrence)
and the scaling behavior in the vicinity of CPs. Nevertheless, as mentioned above, entanglement is not sufficient to account for all the correlation
contained in quantum systems, so this motivates us to apply other correlation witnesses to study their dynamic behaviors combining with QRG method.
To serve as a further comparison, the violation of Bell inequalities \cite{Bell1964,CHSH1969} is also taken into consideration.

The outline of this paper is as follows. In Sec. II and Sec. III, we investigate one dimensional anisotropic XXZ model and XY model respectively, using
several kinds of correlation indicators under the method of QRG. In Sec. IV, we discuss the scaling behavior of these quantifiers when close to CPs.
Sec. V is devoted to the discussion and conclusion. Finally, some technical points are clarified in the Appendix.

\section{CORRELATION ANALYSIS IN THE ANISOTROPIC XXZ MODEL}
First, we recall the QRG method and its application in one dimensional anisotropic XXZ model. In fact, renormalization group refers to a mathematical tool
that allows systematic investigation of the changes of a physical system as viewed at different distance scales. The key point to QRG scheme lies in
reducing the effective degrees of freedom of the system through a recursive procedure until a mathematically tractable situation is reached.
Following Kadanoff's approach (the "block-spin" renormalization group), the (one dimensional) lattice is split into blocks. The Hamiltonian of each
block is diagonalized exactly to obtain the low-lying eigenstates (project operator) to construct the basis for renormalized Hilbert space.
Finally the full Hamiltonian is projected onto the renormalized space to achieve an effective Hamiltonian $H^{eff}$. Here we can
summarize the QRG method as the following steps:

1) Decomposing the Hamiltonian into the intrablock and interblock parts: $H=H^B+H^{BB}$, where $H^B$ is the block Hamiltonian, and
the interblock interaction is denoted as $H^{BB}$.

2) Diagonalization of $H^B$: this procedure is aiming to obtain the low-lying eigenstates and build the projection operator $P_0$ onto the
the low energy subspace.

3) Renormalization of $H^B$ and $H^{BB}$: by virtue of perturbative expansion (see Ref. [30]), the effective (renormalized) Hamiltonian up to the
first-order correction is $H^{eff}=H^{eff}_0+H^{eff}_1$, where $H^{eff}_0=P_0H^BP_0$ and $H^{eff}_1=P_0H^{BB}P_0$.

4) Iteration: repeat $1)\Rightarrow 3)$ to arrive at the final manageable situation.
For more details, we refer the readers to Ref. [30-32].

Kargarian et al. introduced the notion of "renormalization of concurrence" \cite{Kargarian2007}, and they found that this notion truly captures the nonanalytic
behavior of the derivative of entanglement (concurrence) close to the critical point. As a warmup, we first review the renormalization of entanglement in
the one-dimensional anisotropic XXZ model \cite{Kargarian2008}. The Hamiltonian of spin 1/2 XXZ model on a periodic chain of N sites is
\begin{eqnarray}
\label{XXZ}
H(J,\Delta)=\frac{J}{4}\sum^N_i\left(\sigma^x_i\sigma^x_{i+1}+\sigma^y_i\sigma^y_{i+1}+\Delta\sigma^z_i\sigma^z_{i+1}\right),
\end{eqnarray}
where $J$ is exchange constant, $\Delta$ is the anisotropy parameter, and $J,\Delta>0$. $\sigma^\alpha_i (\alpha=x,y,z)$ are standard Pauli matrices
at site $i$. This model is known to be exactly solvable by Bethe Ansatz and critical (gapless) while $0\leq\Delta\leq1$. The Ising regime is $\Delta>1$
and a maximum of concurrence can be reached between two nearest-neighbouring spins at the transition point $\Delta=1$ \cite{Gu2003,Gu2005}.

To construct a renormalized form for the Hamiltonian (\ref{XXZ}), we shall choose a decomposition of three-site blocks. Note that this is requisite in
the sense that it is a guarantee of self-similarity after each iterative step. Ref. \cite{Kargarian2008} gives the degenerate ground states of the block
Hamiltonian as follows
\begin{eqnarray}
|\phi_0\rangle=\frac{1}{\sqrt{2+q^2}}(|\uparrow\uparrow\downarrow\rangle+q|\uparrow\downarrow\uparrow\rangle+|\downarrow\uparrow\uparrow\rangle),
\label{ground1}\\
|\phi'_0\rangle=\frac{1}{\sqrt{2+q^2}}(|\uparrow\downarrow\downarrow\rangle+q|\downarrow\uparrow\downarrow\rangle+|\downarrow\downarrow\uparrow\rangle),
\end{eqnarray}
where $|\uparrow\rangle, |\downarrow\rangle$ are the eigenstates of $\sigma_z$ and
\begin{eqnarray}
q=-\frac{1}{2}(\Delta+\sqrt{\Delta^2+8}),
\end{eqnarray}
Then the effective Hamiltonian of the renormalized chain can be cast into the form
\begin{eqnarray}
H^{eff}=\frac{J'}{4}\sum^{N/3}_i\left(\sigma^x_i\sigma^x_{i+1}+\sigma^y_i\sigma^y_{i+1}+\Delta'\sigma^z_i\sigma^z_{i+1}\right),
\end{eqnarray}
where the iterative relationship is
\begin{eqnarray}
J'=J(\frac{2q}{2+q^2})^2,\, \Delta'=\Delta\frac{q^2}{4}.
\end{eqnarray}
The most important information given in the QRG method are its fixed points. By solving equation $\Delta'=\Delta$, we obtain the trivial
fixed point $\Delta=0$ and also the nontrivial fixed point $\Delta=1$. It is worth noticing that, as was stated previously, XXZ model is critical
for all $0\leq\Delta\leq1$ while QRG method only indicates the single points. Indeed, if appropriate boundary terms are implemented in the
QRG method, then it predicts correctly a line of critical models in the range $0\leq\Delta\leq1$ \cite{Martin1996b}.

In order to calculate quantum discord and other correlation quantities, we consider one of the degenerate ground states. Correspondingly,
the density matrix is defined as
\begin{eqnarray}
\rho_{123}=|\phi_0\rangle\langle\phi_0|,
\end{eqnarray}
with $|\phi_0\rangle$ referring to Eq. (\ref{ground1}) (choosing $|\phi'_0\rangle$ will yield the same results). Since we are focusing on
pairwise correlation functions, without loss of generality, we trace over site 2 to obtain the reduced density matrix between site 1 and 3
\begin{equation}
\label{matrix1}
\rho_{13}=\frac{1}{2+q^2}
\left(\begin{array}{cccc}
q^2 & 0 & 0 & 0 \\
0 & 1 & 1 & 0 \\
0 & 1 & 1 & 0 \\
0 & 0 & 0 & 0
\end{array}\right),
\end{equation}
It is straightforward to compute the concurrence \cite{Wootters1998} of $\rho_{13}$
\begin{eqnarray}
C_{13}=\frac{2}{2+q^2}.
\end{eqnarray}
It is shown that the concurrence between two blocks exhibits
an explicit signature of the quantum phase transitions
at $\Delta=1$. Meanwhile, we notice that some other entanglement measures in the literature,
such as von Neumann entropy or the averaged bipartite entanglement, are shown to be
good indicators of the quantum phase transition.
However, to our best knowledge, there is
no study of analyzing other correlation witnesses beyond entanglement in QRG
framework. In the below section, we will analytically derive
these quantities in detail to see whether they can be
proved really helpful in predicting critical phenomenon.

\subsection{Quantum discord and measure-induced disturbance}
Quantum discord is introduced in Ref. \cite{Ollivier2001} aiming to characterize all the nonclassical correlations present in a
bipartite state. It originates from the inequivalence of two expressions of mutual information in the quantum realm.
Consider a composite bipartite system $\rho^{AB}$, the quantum mutual information is defined as
\begin{eqnarray}
\mathcal{I}(\rho^{AB}):=S(\rho^{A})+S(\rho^{B})-S(\rho^{AB}),
\end{eqnarray}
where $S(\rho)=-Tr(\rho\log_2\rho)$ is the von Neumann entropy, and $\rho^{A(B)}=Tr_{B(A)}(\rho^{AB})$ denote the reduced density operator
of subsystem A(B). On the other hand, if a complete set of von Neumann measurements $\{\Pi^{A}_{k}\}$ (or more generally, POVMs) performed on subsystem A,
an alternative version of quantum mutual information conditioned on this measurement yields
\begin{align}
\mathcal{I}(\rho|\{\Pi^{A}_{k}\}):&=S(\rho^{B})-S(\rho|\{\Pi^{A}_{k}\}),\\
&=S(\rho^{B})-\sum_kp_kS(\rho^B_k),
\end{align}
with $p_k=Tr(\Pi^{A}_{k}\rho^{AB})$ and $\rho^B_k=Tr_A(\Pi^{A}_{k}\rho^{AB})/p_k$.
To eliminate the dependence on specific measurement, one takes the optimization procedure to obtain
\begin{eqnarray}
\mathcal{J}(\rho):=\max_{\{\Pi^{A}_{k}\}}\mathcal{I}(\rho|\{\Pi^{A}_{k}\}),
\end{eqnarray}
which has been suggested by Henderson and Vedral \cite{Henderson2001} as a measure to quantify the purely classical part of correlations.
The discrepancy between the original quantum mutual information $\mathcal{I}$ and the measurement-induced quantum mutual information $\mathcal{J}$
is defined as the so called quantum discord
\begin{align}
\mathcal{D}_A(\rho):&=\mathcal{I}(\rho)-\mathcal{J}(\rho),\\
&=S(\rho^A)-S(\rho^{AB})+\min_{\{\Pi^{A}_{k}\}}\sum_kp_kS(\rho^B_k).
\end{align}

Now we are going to deal with the situation that we come across. The density matrix defined in Eq. (\ref{matrix1}) is a two-qubit X-shaped state.
A general X-state have the visual appearance
\begin{equation}
\label{X}
\rho^{\chi}=
\left(\begin{array}{cccc}
\rho_{11} & 0 & 0 & \rho_{14} \\
0 & \rho_{22} & \rho_{23} & 0 \\
0 & \rho_{23}^\ast & \rho_{33} & 0 \\
\rho_{14}^\ast & 0 & 0 & \rho_{44}
\end{array}\right),
\end{equation}
which has seven real parameters. However, up to local unitary equivalence, we can assume $\rho_{14} $ and $\rho_{23}$ are also real and
in fact there are only five independent parameters (note that QD is invariant under local unitary transformations). Alternatively,
if we represent the X-state in Bloch decomposition, then the five characterizing parameters can be expressed as
\begin{eqnarray}
\label{parameter}
x&=&Tr(\sigma^A_z\rho^\chi)=\rho_{11}+\rho_{22}-\rho_{33}-\rho_{44},\nonumber\\
y&=&Tr(\sigma^B_z\rho^\chi)=\rho_{11}-\rho_{22}+\rho_{33}-\rho_{44},\nonumber\\
t_1&=&Tr(\sigma^A_x\sigma^B_x\rho^\chi)=2\rho_{14}+2\rho_{23},\\
t_2&=&Tr(\sigma^A_y\sigma^B_y\rho^\chi)=-2\rho_{14}+2\rho_{23},\nonumber\\
t_3&=&Tr(\sigma^A_z\sigma^B_z\rho^\chi)=\rho_{11}-\rho_{22}-\rho_{33}+\rho_{44},\nonumber
\end{eqnarray}
Except for some numerical evaluations for a restricted subset of two-qubit X states \cite{Maziero2010,Fanchini2010}, Ref. \cite{Ali2010} has presented a
algorithm to calculate QD for general two-qubit X states: the optimal measurement is in a universal finite set $\{\sigma_x,\sigma_y,\sigma_z\}$ (see also Ref. \cite{Chen2011,Yu2011}).
Nonetheless, a counterexample is given by Ref. \cite{Lu2011} to disprove the algorithm and this fact elucidates the state-dependence in the optimization
for general X states. Furthermore, recent progress toward this problem has been made by Ref. \cite{Chen2011,Yu2011}, which identifies
a large class of X states whose QD can be derived analytically from the measurement strategy described above.

Keeping these technical preparations in mind, let's turn to density matrix (\ref{matrix1}), the spectrum of which is
$\lambda(\rho_{13})=\{0,0,\frac{2}{2+q^2},\frac{q^2}{2+q^2}\}$. First we note that $\rho_{22}=\rho_{33}$ and then
\begin{equation}
S(\rho^A)=S(\rho^B)=\frac{1}{2+q^2}
\left(\begin{array}{cc}
1+q^2 & 0 \\
0 & 1
\end{array}\right),
\end{equation}
which means $\mathcal{D}_A(\rho)=\mathcal{D}_B(\rho)$ \cite{Henderson2001} (here A and B denotes site 1 and 3 respectively). According to the Theorem
in Ref. \cite{Chen2011}, it is easy to check that the optimal observable for state (\ref{matrix1}) is not $\sigma_z$ but $\sigma_x$. For the reason
that will be clear later, we also gives discord measured by $\sigma_z$ (for brevity we put the general formulas in the Appendix) which is actually equal
to the concurrence
\begin{eqnarray}
\mathcal{D}_A^{\sigma_z}=\frac{2}{2+q^2},
\end{eqnarray}
When the optimal measurement $\{\frac{1}{2}(I\pm\sigma^A_x)\}$ is performed on A, we directly acquire the QD of state (\ref{matrix1})
\begin{align}
\mathcal{D}_A^{\sigma_x}=&-\frac{1+q^2}{2+q^2}\log_2(\frac{1+q^2}{2+q^2})-\frac{1}{2+q^2}\log_2(\frac{1}{2+q^2})\nonumber\\
&+\frac{2}{2+q^2}\log_2(\frac{2}{2+q^2})+\frac{q^2}{2+q^2}\log_2(\frac{q^2}{2+q^2})\nonumber\\
&+f\left[\frac{4+q^4}{(2+q^2)^2}\right],
\end{align}
where $f(z):=-\frac{1+\sqrt{z}}{2}\log_2\frac{1+\sqrt{z}}{2}-\frac{1-\sqrt{z}}{2}\log_2\frac{1-\sqrt{z}}{2}$.
Numerical evaluation shows $\mathcal{D}_A=\mathcal{D}_A^{\sigma_x}$ is strictly less than $\mathcal{D}_A^{\sigma_z}$, as we expect.

Based on the definition introduced by Luo \cite{Luo2008a}, measure-induced disturbance (MID) is defined as the difference of quantum mutual
information before and after measurement
\begin{eqnarray}
MID(\rho_{AB})=I(\rho_{AB})-I(\Pi(\rho_{AB})),
\end{eqnarray}
The measurement $\Pi=\{\Pi^A_i\otimes\Pi^B_j\}$ is induced by the spectral decompositions of the reduced states, $\rho^A=\Sigma_ip_i\rho^A_i$
and $\rho^B=\Sigma_ip_i\rho^B_i$, which leaves the marginal information invariant. By taking $\Pi^A_i=|i\rangle\langle i|$ and
$\Pi^B_j=|j\rangle\langle j|$ (which is unique in our case), we have $\Pi(\rho_{AB})=\frac{1}{2+q^2}diag\{q^2,1,1,0\}$ and consequently
\begin{eqnarray}
MID=S(\Pi(\rho_{AB}))-S(\rho_{AB})=\frac{2}{2+q^2},
\end{eqnarray}
So MID of state (\ref{matrix1}) coincides with $\mathcal{D}_A^{\sigma_z}$ and concurrence. However, this is not a coincidence.
In the Appendix we will prove that MID is exactly equal to the QD measured by $\sigma_z$ for general X-structured states.

\begin{figure}[htbp]
\begin{center}
\includegraphics[width=.40\textwidth]{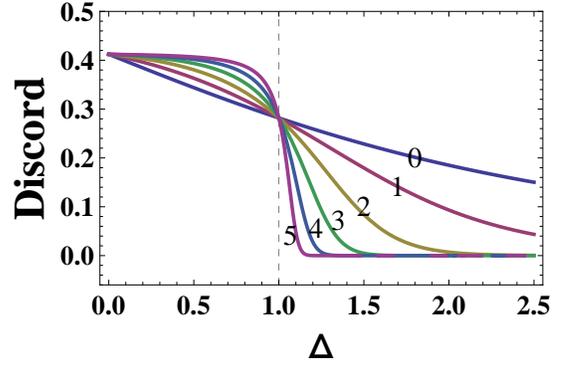} {}
\end{center}
\caption{(Color online) The evolution of the quantum discord versus $\Delta$
in terms of QRG iteration steps in the XXZ model.
}\label{QD1}
\end{figure}

In Figure \ref{QD1}, we illustrate the evolution of QD versus $\Delta$ for different QRG steps. Notice that the iterative relationship
we adopt here and later in the calculation is Eq. (4) and (6). The plots of QD cross each other
at the critical point $\Delta=1$. In comparison with concurrence demonstrated in Ref. \cite{Kargarian2008}, QD also develops
two saturated values, which are associated with the two different phases, spin-liquid and N\'{e}el phases. Note that after
enough iteration steps for $0\leq\Delta<1$, $D_A\approx0.412154<C=0.5$ while for $\Delta>1$, $D_A\rightarrow0$. In addition,
since MID is exactly equal to concurrence, MID can obviously exhibit a QPT at $\Delta=1$.

\subsection{Geometric discord and measure-induced nonlocality}
In this subsection we calculate GD and MIN. Recently, Daki\'{c} et al. introduced the following geometric measure of quantum discord
based on the Hilbert-Schmidt norm \cite{Dakic2010}
\begin{eqnarray}
\mathcal{D}^{G}_A(\rho):=\min_{\chi\in\Omega}\|\rho-\chi\|^2,
%\label{}
\end{eqnarray}
where $\Omega$ denotes the set of zero-discord states and $\|\rho-\chi\|^2=Tr(\rho-\chi)^2$ is the square of
Hilbert-Schmidt norm. For two-qubit case, a closed form of expression for geometric discord can be achieved
\begin{eqnarray}
\label{QD}
\mathcal{D}^{G}_A(\rho)=\frac{1}{4}(\|\vec{x}\|^2+\|T\|^2-\lambda_{max}),
\end{eqnarray}
where $x_i=Tr(\sigma^A_i\rho)$ are components of the local Bloch vector for subsystem A, $T_{ij}=Tr(\sigma^A_i\sigma^B_j\rho)$
are components of the correlation matrix, and $\vec{x}:=(x_1,x_2,x_3)^t$, $T:=(T_{ij})$, $\lambda_{max}$ is the largest eigenvalue of
the matrix $K=\vec{x}\vec{x}^t+TT^t$ (here the superscript t denotes transpose). It is worth emphasizing that, Luo and Fu presented
an equivalent but simplified version of the geometric discord \cite{Luo2010a}
\begin{eqnarray}
\mathcal{D}^{G}_A(\rho)=\min_{\Pi^A}||\rho-\Pi^A(\rho)||^2,
\end{eqnarray}
where the minimum is over all von Neumann measurements $\Pi^A=\{\Pi^A_k\}$ on subsystem A. Intuitively, in some sense dual to GD,
another measure quantifying nonlocal effect caused by locally invariant measurements was introduced Luo and Fu \cite{Luo2011}
\begin{eqnarray}
MIN_A(\rho)=\max_{\Pi^A}||\rho-\Pi^A(\rho)||^2,
\end{eqnarray}
with an extra constraint that von Neumann measurements $\Pi^A=\{\Pi^A_k\}$ do not disturb $\rho^A$ locally, which means
$\rho^A=\sum_k\Pi_k^A\rho^A\Pi_k^A$.

\begin{figure}[htbp]
\begin{center}
\includegraphics[width=.40\textwidth]{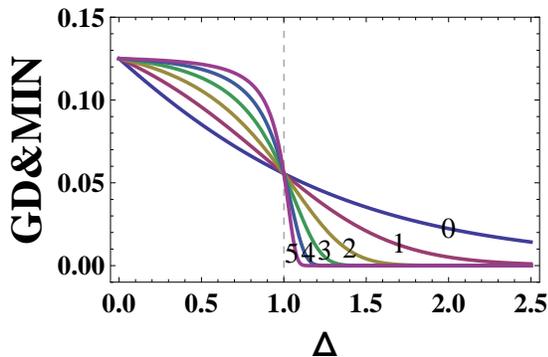} {}
\end{center}
\caption{(Color online) The evolution of the geometric discord and measure-induced nonlocality versus $\Delta$
in terms of QRG iteration steps in the XXZ model.
}\label{GD1}
\end{figure}

For X state and its characterizing parameters defined in Eqs. (\ref{parameter}), we have $\vec{x}=(0,0,x)^t$ and
$T=diag\{t_1,t_2,t_3\}$ and GD of X state reads
\begin{align}
\mathcal{D}^{G}_A(\rho^\chi)=&\frac{1}{4}(t_1^2+t_2^2+t_3^2+x^2\nonumber\\
&-\max\{t_1^2,t_2^2,t_3^2+x^2\}),
\label{GDX}
\end{align}
Besides, using Theorem 3 in Ref. \cite{Luo2011}, MIN can be obtained for two-qubit X states
\begin{eqnarray}
MIN_A(\rho^\chi)=\left\{\begin{array}{cc}
\frac{1}{4}(t_1^2+t_2^2), & \mbox{ if } \, x\neq0\\
\frac{1}{4}(t_1^2+t_2^2+t_3^2-\lambda_{min}), & \mbox{ if } \, x=0
\end{array}\right.
\label{MINX}
\end{eqnarray}
with $\lambda_{min}=\min\{t_1^2,t_2^2,t_3^2\}$. Applying these formulas to the state (\ref{matrix1}), we obtain
\begin{eqnarray}
\mathcal{D}^{G}_A=MIN_A=\frac{1}{4}(t_1^2+t_2^2)=\frac{2}{(2+q^2)^2}=\frac{1}{2}C^2,
\end{eqnarray}
by noting that $t_3^2+x^3=(\frac{q^2-2}{2+q^2})^2+(\frac{q^2}{2+q^2})^2\geq(\frac{2}{2+q^2})^2=t_1^2=t_2^2$ and $x\neq0$
since $|q|\geq\sqrt{2}$ for $\Delta\geq0$. The variation of GD and MIN versus $\Delta$ has been plotted in Figure \ref{GD1}.
It is no surprise that they can indicate the precise location of the critical point $\Delta=1$ since $\mathcal{D}^{G}_A=MIN_A=\frac{1}{2}C^2$.

\subsection{Bell violation}
Quantum nonlocality, as revealed by the violation of Bell-type inequalities, refers to many-system measurement correlations that cannot
be simulated by any local hidden variable theory. In particular, for two-qubit pure states, the presence of entanglement guarantees violation
of a Bell inequality (Gisin's Theorem) \cite{Gisin1991}. However, for mixed stares the situation becomes more complicated \cite{Werner1989}.
Here we restrict ourself to the CHSH inequality. The Bell operator corresponding to CHSH inequality can be formulated in the following form
\begin{eqnarray}
\mathcal{B}_{CHSH}=\bm{a}\cdot\bm{\sigma}\otimes(\bm{b}+\bm{b'})\cdot\bm{\sigma}
+\bm{a'}\cdot\bm{\sigma}\otimes(\bm{b}-\bm{b'})\cdot\bm{\sigma},
\end{eqnarray}
where $\bm{a}$, $\bm{a'}$, $\bm{b}$, $\bm{b'}$ are the unit vectors in $\mathbb{R}^3$, and $\bm{\sigma}=(\sigma_x,\sigma_y,\sigma_z)$.
Then the well-known CHSH inequality is expressed as
\begin{eqnarray}
%\label{}
B=|\langle\mathcal{B}_{CHSH}\rangle_\rho|=|Tr(\rho\mathcal{B}_{CHSH})|\leq2.
\end{eqnarray}
According to the Horodecki criterion \cite{Horodecki1995}, the maximum violation of CHSH
inequality is given by
\begin{eqnarray}
B_{CHSH}^{max}&=&\max_{\bm{a}, \bm{a'}, \bm{b}, \bm{b'}}Tr(\rho\mathcal{B}_{CHSH}),\nonumber\\
&=&2\sqrt{\max_{i<j}(u_i+u_j)},
\end{eqnarray}
where $u_i$, $i=1,2,3$ are the eigenvalues of $U=T^{t}T$.

As to X states, the matrix $T$ is diagonal and $T^{t}T=diag\{t_1^2,t_2^2,t_3^2\}$. Therefore the maximal violation of the
CHSH inequality for X states can be simplified to
\begin{eqnarray}
B_{CHSH}^{max}(\rho^\chi)=2\sqrt{\sum_{i=1}^3t_i^2-\lambda_{min}},
\label{BellX}
\end{eqnarray}
with $\lambda_{min}=\min\{t_1^2,t_2^2,t_3^2\}$. Thus, the (maximal) Bell violation of state (\ref{matrix1}) is obtained
\begin{eqnarray}
B_{max}=2\sqrt{\max\left\{\frac{8}{(2+q^2)^2},\frac{(q^2-2)^2+4}{(2+q^2)^2}\right\}}.
\end{eqnarray}

\begin{figure}[htbp]
\begin{center}
\includegraphics[width=.40\textwidth]{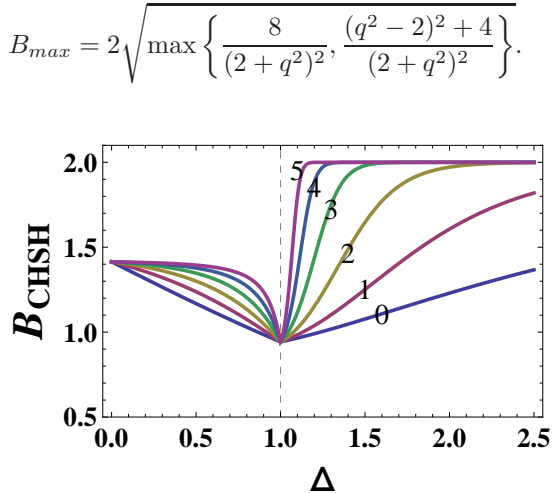} {}
\end{center}
\caption{(Color online) The evolution of the violation of CHSH inequality versus $\Delta$
in terms of QRG iteration steps in the XXZ model.
}\label{Bell1}
\end{figure}

The value of maximum Bell violation under QRG iterations is displayed in Figure \ref{Bell1}. Interestingly, we can observe that
the block-block correlations never violate the CHSH inequality, but still evidently exhibit a QPT. At the critical point $\Delta=1$,
the (maximum) Bell violation is a fixed nonzero constant $B_c=2\sqrt{2}/3\approx0.942809$, irrespective of the iterative steps.
What's more, beyond the critical point, Bell violation also develops two saturated values: one is $B=\sqrt{2}$ for $0\leq\Delta<1$
and one is $B=2$ for $\Delta>1$, which is in sharp contrast to the behavior of concurrence.

\section{CORRELATION ANALYSIS IN THE ANISOTROPIC XY MODEL}
In this section we embark on studying the relationship between the QPT and quantum correlation witnesses in the spin-1/2 XY model with
the QRG method. The Hamiltonian of the XY model on a periodic chain with N sites reads
\begin{eqnarray}
\label{XY}
H(J,\gamma)=\frac{J}{4}\sum^N_i\left[(1+\gamma)\sigma^x_i\sigma^x_{i+1}+(1-\gamma)\sigma^y_i\sigma^y_{i+1}\right],
\end{eqnarray}
where $J$ is exchange coupling constant, $\gamma$ is the anisotropy parameter. The XY model reduces to the XX model for $\gamma=0$
or the Ising model for $\gamma=1$. In the parameter range $0<\gamma\leq1$, it falls into the Ising universality class.
To implement the QRG method, we still choose three sites as a block. In Ref. \cite{Ma2011a} Ma et al. obtained two doubly degenerate
eigenvalues of the block Hamiltonian as below
\begin{align}
|\Phi_0\rangle=&\frac{1}{2\sqrt{1+\gamma^2}}(-\sqrt{1+\gamma^2}|\uparrow\uparrow\downarrow\rangle+\sqrt{2}|\uparrow\downarrow\uparrow\rangle\nonumber\\
&-\sqrt{1+\gamma^2}|\downarrow\uparrow\uparrow\rangle+\sqrt{2}\gamma|\downarrow\downarrow\downarrow\rangle),\label{ground2}\\
|\Phi'_0\rangle=&\frac{1}{2\sqrt{1+\gamma^2}}(-\sqrt{2}\gamma|\uparrow\uparrow\uparrow\rangle+\sqrt{1+\gamma^2}|\downarrow\uparrow\uparrow\rangle\nonumber\\
&-\sqrt{2}|\downarrow\uparrow\downarrow\rangle+\sqrt{1+\gamma^2}|\downarrow\downarrow\uparrow\rangle),
\end{align}
After projection onto the renormalized subspace, the effective Hamiltonian can be written as
\begin{eqnarray}
%\label{}
H^{eff}=\frac{J'}{4}\sum^N_i\left[(1+\gamma')\sigma^x_i\sigma^x_{i+1}+(1-\gamma')\sigma^y_i\sigma^y_{i+1}\right],
\end{eqnarray}
with the iterative relationship
\begin{eqnarray}
J'=J\frac{3\gamma^2+1}{2(1+\gamma^2)},\, \gamma'=\frac{\gamma^3+3\gamma}{3\gamma^2+1}.
\end{eqnarray}
Naturally we are most concerned with the CP information. The stable and unstable fixed points can be gotten by solving $\gamma'=\gamma$.
The stable fixed points locate at $\gamma=\pm1$, and the unstable fixed point is $\gamma=0$ which separates
the spin-fluid phase ($\gamma=0$) from the the N\'{e}el phase ($0<|\gamma|\leq1$).

Similarly, we consider one of the degeneracy ground states to construct the pure-state density matrix
\begin{eqnarray}
\varrho_{123}=|\Phi_0\rangle\langle\Phi_0|,
\end{eqnarray}
The result of choosing $|\Phi'_0\rangle$ will be the same. By tracing out site 2, we arrive at the reduced density matrix
\begin{equation}
\label{matrix2}
\varrho_{13}=\frac{1}{4(\gamma^2+1)}
\left(\begin{array}{cccc}
2 & 0 & 0 & 2\gamma \\
0 & \gamma^2+1 & \gamma^2+1 & 0 \\
0 & \gamma^2+1 & \gamma^2+1 & 0 \\
2\gamma & 0 & 0 & 2\gamma^2
\end{array}\right),
\end{equation}
The concurrence between the sites 1 and 3 is given as
\begin{eqnarray}
C_{13}=\frac{1}{2}-\frac{|\gamma|}{1+\gamma^2}.
\end{eqnarray}

\subsection{Quantum discord and measure-induced disturbance}
Before calculating QD and other correlation quantities, we regard all these measures as a function of $\rm{g}$, where
\begin{eqnarray}
\rm{g}=\frac{1+\gamma}{1-\gamma},
\end{eqnarray}
The reason is twofold: bringing in such a variable is not only convenient for us to compare the results with that of \cite{Ma2011a}, but also
useful in the derivation process. According to Ref. \cite{Chen2011,Yu2011}, it can be verified that $\sigma_z$ is not the optimal observable
and the choose of optimal observable depends on the value of $\gamma$, or more accurately, the sign of $\gamma$: the the optimal observable
to measure is $\sigma_x$ if $\gamma\geq0$ (which is equivalent to $t_1\geq t_2$ or $|\rm{g}|\geq1$), and $\sigma_y$ if $\gamma<0$ ($|\rm{g}|<1$).
In spite of this fact, we still provide the discord measured by $\sigma_z$ here
\begin{align}
\mathcal{D}^{\sigma_z}=&-\frac{1}{2(\gamma^2+1)}\log_2\frac{1}{2(\gamma^2+1)}\nonumber\\
&-\frac{\gamma^2}{2(\gamma^2+1)}\log_2\frac{\gamma^2}{2(\gamma^2+1)},
\end{align}
We again remark that $\varrho_{22}=\varrho_{33}$ for the density matrix (\ref{matrix2}) and thus
\begin{equation}
S(\rho^A)=S(\rho^B)=\frac{1}{4(\gamma^2+1)}
\left(\begin{array}{cc}
\gamma^2+3 & 0 \\
0 & 3\gamma^2+1
\end{array}\right),
\end{equation}
So we do not need to specify on which subsystem the measurement is performed. The spectrum of (\ref{matrix2}) is $\lambda(\varrho)=\{1/2,1/2,0,0\}$,
irrespective of the value of $\gamma$. After some lengthy but standard algebra, one finally gets
\begin{align}
\mathcal{D}=&S(\varrho^A)-S(\varrho^{AB})+f(y^2+\max\{t^2_{1,2}\})\nonumber\\
=&-\frac{\gamma^2+3}{4(\gamma^2+1)}\log_2\frac{\gamma^2+3}{4(\gamma^2+1)}\nonumber\\
&-\frac{3\gamma^2+1}{4(\gamma^2+1)}\log_2\frac{3\gamma^2+1}{4(\gamma^2+1)}\nonumber\\
&-1+f\left[\frac{(|\gamma|+1)^2}{2(\gamma^2+1)}\right],
\end{align}
with $y,t_1,t_2$ and function $f(\cdot)$ defined the same as above.

\begin{figure}[htbp]
\begin{center}
\includegraphics[width=0.4\textwidth ]{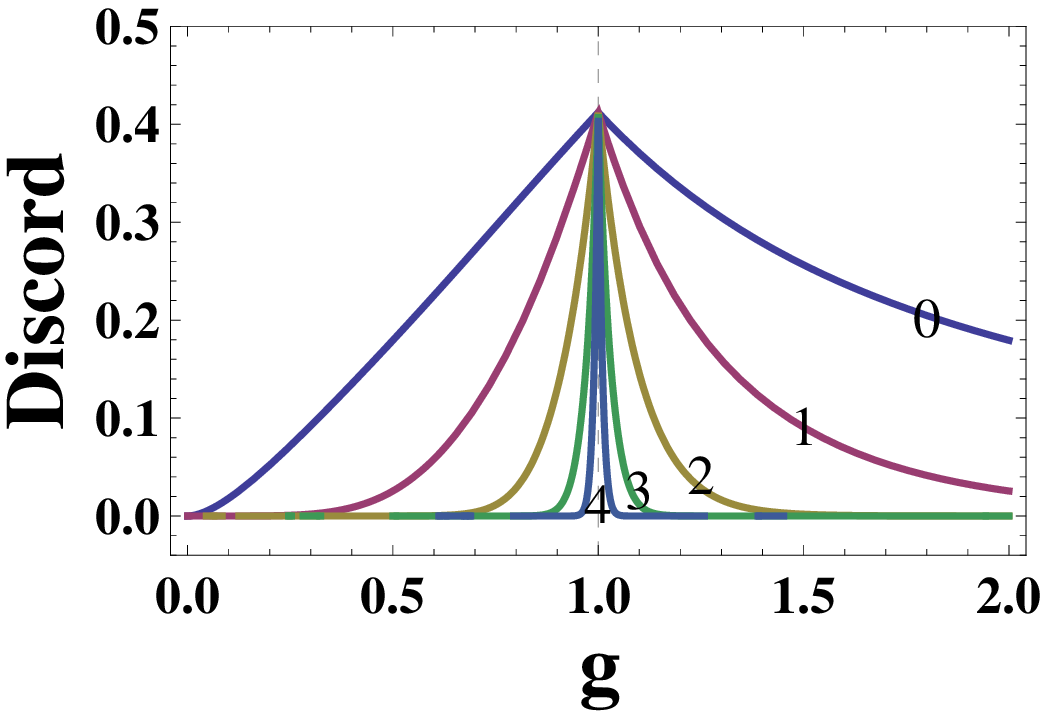}\\
\includegraphics[width=0.4\textwidth ]{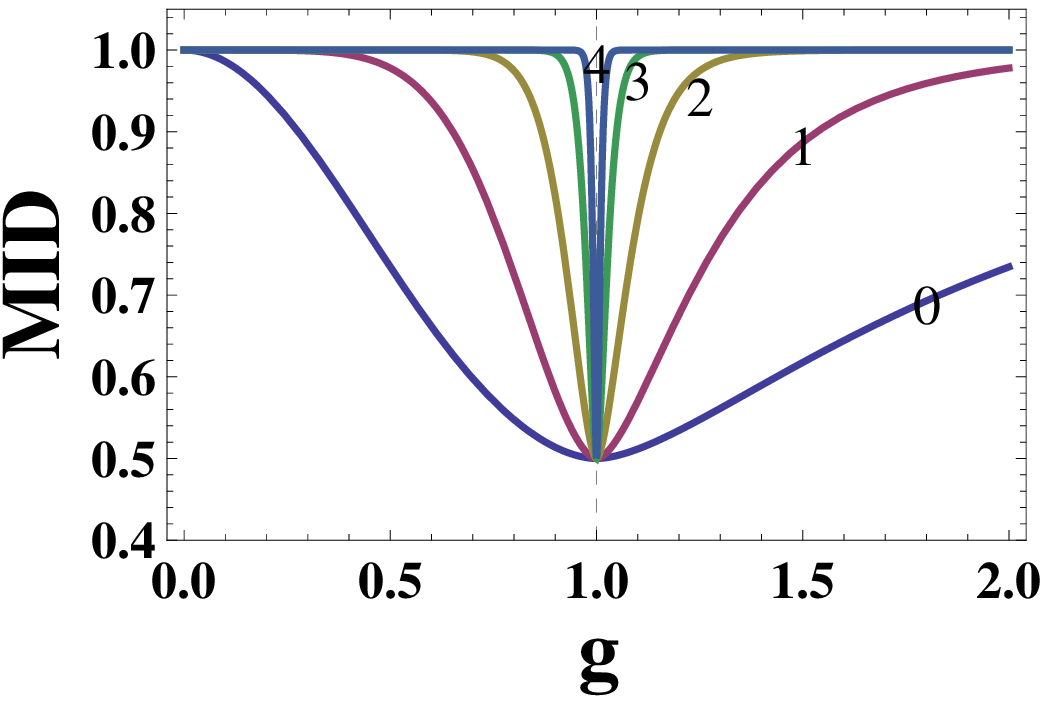}
\end{center}
\caption{(Color online) The evolution of quantum discord and measure-induced disturbance versus $\Delta$
in terms of QRG iteration steps in the XY model.
}\label{QD2}
\end{figure}

From the proof in the Appendix, we already know that MID is exactly equal to $\mathcal{D}^{\sigma_z}$.
The QD and MID between site 1 and 3 have been plotted in Figure \ref{QD2}
(recall that for this model the iterative relationship is just Eq. (39)). The dynamic behavior of QD
in each iteration step is analogous to that of concurrence, but not totally the same. At the critical
point $\rm{g}=1$, QD reaches a nonzero constant $\mathcal{D}\approx0.412154$, which once again indicates that spin-fluid
phase contains quantum correlations as already shown in XXZ model. In contrast to QD, MID also
shows the nonanalytic property, however, for $0\leq\rm{g}<1$ and $\rm{g}>1$ MID does not fall to zero but
gets to another nonzero constant $MID=1$.

\subsection{Geometric discord, measure-induced nonlocality and Bell violation}
\begin{figure}[htbp]
\begin{center}
\includegraphics[width=.40\textwidth]{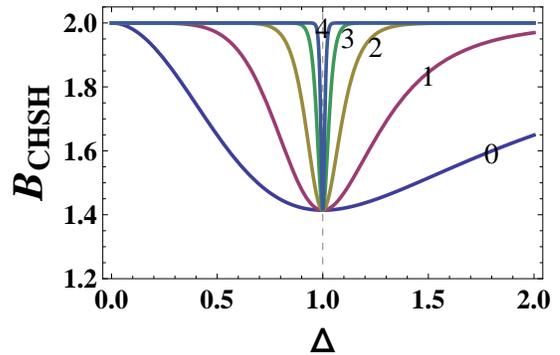} {}
\end{center}
\caption{(Color online) The evolution of the violation of CHSH inequality versus $\Delta$
in terms of QRG iteration steps in the XY model.
}\label{Bell2}
\end{figure}

In section II we have obtained the analytic formulas for GD, MIN and Bell violation. Employing Eqs. (\ref{GDX}), (\ref{MINX}) and (\ref{BellX}),
the GD, MIN and Bell violation for the state (\ref{matrix2}) are listed as follows (in the derivation note that $\gamma\geq0$ and $\gamma<0$ correspond to
$|\rm{g}|\geq1$ and $|\rm{g}|<1$ respectively)
\begin{eqnarray}
\mathcal{D}^{G}&=&\frac{1}{4}\left(\frac{1}{2}-\frac{|\gamma|}{1+\gamma^2}\right)=\frac{1}{4}C,\\
MIN&=&\frac{\gamma^4+6\gamma^2+1}{8(\gamma^2+1)^2},\\
B_{max}&=&\frac{\sqrt{2\gamma^4+12\gamma^2+2}}{\gamma^2+1}=4\sqrt{MIN}.
\end{eqnarray}
Here the Bell violation versus $\rm{g}$ changing for different iterations is depicted in Figure \ref{Bell2}.
It is clearly seen that in XY model the block-block correlation is still not strong enough to violation
the CHSH inequality although Bell violation displays the nonanalytic behavior at $\rm{g}=1$ ($\gamma=0$).

\section{NONANALYTIC AND SCALING BEHAVIOR}
So far, we employ the QRG method to investigate the
\textit{block-block correlations} of one-dimensional XXZ and XY spin
models. As we have described in the QRG approach,
the size of a large system ($N=3^{n+1}$) can
be effectively rescaled to three sites with the renormalized couplings of the $n$th RG
iteration. Therefore, the quantum correlations between the two
renormalized sites represents the correlations between two
parts of the system each containing $N/3$ sites effectively. In
this sense, we can refer to these quantities considered in this work
as block-block correlations. It has been demonstrated that the first derivative of all
these correlation measures shows a nonanalytic behavior in the vicinity of the critical point.
Furthermore, the scaling property of entanglement has also been observed in Ref. \cite{Kargarian2008,Ma2011a}, which
is related to the divergence of the correlation length as the critical point is approached.
\begin{figure}[htbp]
\begin{center}
\includegraphics[width=.40\textwidth]{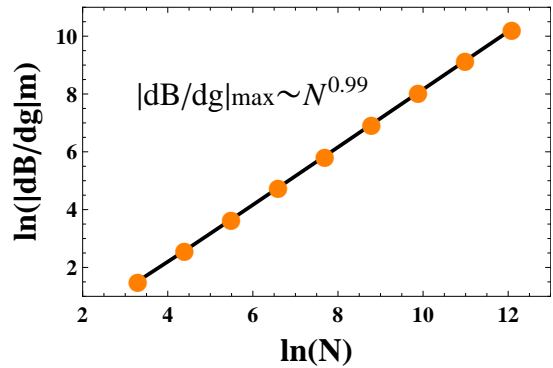} {}
\end{center}
\caption{(Color online) The logarithm of the absolute value of minimum, $\ln(|dB/d\rm{g}|_m)$, versus the logarithm of chain size,
$\ln(N)$, which is linear and displays a scaling behavior ($B$ is the Bell violation in the XY model).
}\label{scale1}
\end{figure}

\begin{figure}[htbp]
\begin{center}
\includegraphics[width=.40\textwidth]{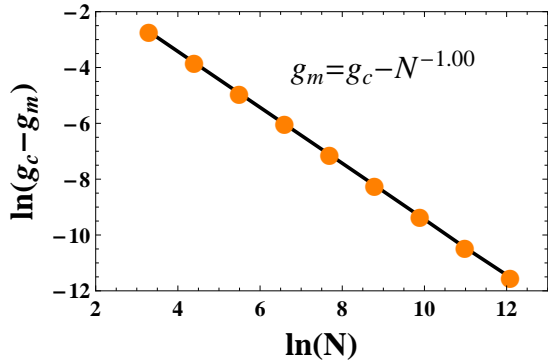} {}
\end{center}
\caption{(Color online) The scaling behavior of $\rm{g}_{max}$ in terms of system size $N$ where $\rm{g}_{max}$
is the position of the minimum derivative of Bell violation in the XY model.
}\label{scale2}
\end{figure}

Aiming to compare with the previous results concerning entanglement, we explicitly show the nonanalytic phenomenon
and scaling behavior of other correlation witnesses. Here, we focus on the dynamic property of nonlocality, that is,
the maximal violation of Bell-CHSH inequality, since our results display that quantum discord behaves more similarly to entanglement
(see Figure \ref{QD1} and \ref{QD2}). First, we have analyzed the scaling behavior of $y=|dB/d\rm{g}|_{\rm{g}_m}$ versus the size of
the system $N$ in the XY model where $\rm{g}_m$ is the position of the minimum of $dB/d\rm{g}$. We have plotted $\ln(y)$ versus $\ln(N)$ in Figure
\ref{scale1} which shows a linear behavior. Numerical calculation tells us that the exponent for this behavior is
$|dB/d\rm{g}|_{\rm{g}_m}\sim N^{0.99}$. As a companion, our analysis also reveals that the position of the minimum $\rm{g}_m$
of $dB/d\rm{g}$ gradually tends to the critical point $\rm{g}_c=1$, as shown in Figure \ref{scale2} (the numerical relation is
$\rm{g}_m=\rm{g}_c-N^{-1.00}$). These results convince us that Bell violation truly signify the criticality of the spin system.

\section{DISCUSSION AND CONCLUSION}
In this paper we have investigated the performance of various correlation measures in quantum phase transition,
exploiting the quantum renormalization group method. In most previous literature along this line, only entanglement
(concurrence) has been utilized as an information-theoretic tool to evaluate the critical properties of the spin systems.
However, it is now well known that entanglement can not account for all the aspects of quantum correlations, which in turn
motivates us to clarify whether other correlations witnesses (including Bell-CHSH violation) are useful in such a
circumstance. Indeed, there are several points that deserves our attention: (i) The quantum discord and other discord-like
measures turn out to be as good as entanglement to detect the quantum phase transition in the anisotropic XXZ and XY models.
Nevertheless, it is apparent that the dynamic processes of these quantities are not totally similar with entanglement
and even when concurrence vanishes there still exists some kind of quantum correlations which is not captured by entanglement.
(ii) Interestingly, our result shows that CHSH inequality can never be violated in the entire iteration procedure,
which indicates block-block entanglement can not revealed by the CHSH inequality. Moreover, the nonanalytic and scaling behaviors
of Bell violation have been justified by numerical calculations.

On the other hand, the two cases handled in this work can be regard as perfect examples to apply the QD algorithm raised in
Ref. \cite{Chen2011,Yu2011}, where the optimal measurements to achieve QD can be exactly determined. Besides,
we are convinced that the whole analysis in this paper can be extended to many other spin models, since the reduced density matrices
are usually highly symmetric and can be casted into X-shaped states \cite{Kargarian2009,Ma2011b,Jafari2008,Langari1998,Jafari2006}.
Very recently, it has been reported that Bell inequality is able to signal QPT and it can never be violated in the corresponding
spin models \cite{Batle2010,Justino2012}. In these works, the nearest-neighboring spin-spin correlation functions are invoked
to compute the Bell violation, which are usually complex and lengthy. However, we resort to the QRG framework and also illustrate
no violation can be discovered in each iteration step, which implies some intrinsic feature of long-scale corrections. The connection between
these observations will be attractive and may need further investigation. Finally, we would like to mention that it might be
interesting to apply the same approach to high-dimensional systems, where a straightforward numerical analysis could be performed
for some measures of quantum correlations.

\begin{acknowledgments}
This work was supported by the National Basic Research Program of China (Grants No. 2011CBA00200 and No. 2011CB921200),
National Natural Science Foundation of China (Grant NO. 60921091), and China Postdoctoral Science Foundation (Grant No. 20100480695).
\end{acknowledgments}
%%%%%%%%%%%%%%%%%%%%%%%%%%%%%%%%%%%%%%%%%%%%%%%%%%%%%%%%%%%%%%%%%%%%%%%%%%%%%%%%%%%%%%%%%%%%%
\appendix
\section{Analytic proof of $MID=\mathcal{D}^{\sigma_z}$ for general X states}
Here, we show that measure-induced disturbance is exactly equal to the quantum discord measured by $\sigma_z$
for general X-structured states. The expressions of QD and MID are formulated as follows
\begin{align}
\mathcal{D}_A(\rho)=S(\rho^A)-S(\rho)+\min_{\{\Pi^{A}_{k}\}}\sum_kp_kS(\rho^B_k),\\
MID(\rho)=I(\rho)-I(\Pi(\rho))=S(\Pi(\rho))-S(\rho),
\end{align}
Note that the two-side measurements $\Pi=\{\Pi^A_i\otimes\Pi^B_j\}$ employing in MID depend on the spectral decompositions of the reduced states.
Since we only consider the discord measured by $\sigma_z$, that is, $\{\frac{1}{2}(I\pm\sigma_z)\}=\{|0\rangle\langle0|,|1\rangle\langle1|\}$,
the conditional states for general X states (\ref{X}) can be obtained
\begin{align}
\rho^B_0=\frac{1}{\rho_{11}+\rho_{22}}
\left(\begin{array}{cc}
\rho_{11} & 0 \\
0 & \rho_{22}
\end{array}\right),\nonumber\\
\rho^B_1=\frac{1}{\rho_{33}+\rho_{44}}
\left(\begin{array}{cc}
\rho_{33} & 0 \\
0 & \rho_{44}
\end{array}\right),
\end{align}
with $p_0=\rho_{11}+\rho_{22}$ and $p_1=\rho_{33}+\rho_{44}$.
The reduced states are all diagonal states
\begin{align}
\rho^A=
\left(\begin{array}{cc}
\rho_{11}+\rho_{22} & 0 \\
0 & \rho_{33}+\rho_{44}
\end{array}\right),\nonumber\\
\rho^B=
\left(\begin{array}{cc}
\rho_{11}+\rho_{33} & 0 \\
0 & \rho_{22}+\rho_{44}
\end{array}\right),
\end{align}
Therefore, we can take $\Pi^A_i=|i\rangle\langle i|$, $\Pi^B_j=|j\rangle\langle j|$ ($i,j=0,1$).
To prove $MID=\mathcal{D}^{\sigma_z}$, all we need is to verify $S(\rho^A)+\sum_kp_kS(\rho^B_k)=S(\Pi(\rho))$.
In fact, it is easy to find
\begin{eqnarray}
S(\rho^A)+\sum_kp_kS(\rho^B_k)&=&-\sum_{ii}\rho_{ii}\log_2(\rho_{ii})\nonumber\\
&=&S(\Pi(\rho))
\end{eqnarray}
with $ii=11,22,33,44$. Moreover, when the measurement $\sigma_z$ is performed on subsystem B, the situation is the same.
To sum up, we arrive at the relationship $\mathcal{D}^{\sigma_z}_A=\mathcal{D}^{\sigma_z}_B=MID$ for X states.

\section{Analytic formula of $\mathcal{D}^{\sigma_x}(\mathcal{D}^{\sigma_y})$ for general X states}
By definition, we need to evaluate the conditional state $\rho^B_k$ and the corresponding probability $p_k$, since
$S(\rho^A)$ and $S(\rho^{AB})$ are easy to compute. Let $\{\Pi_{k}^A=\frac{1}{2}(I\pm\sigma_x)\}$ ($k=\pm$)
be the local measurement for subsystem A, then
\begin{align}
\rho_{\pm}^B&=\frac{1}{p_{\pm}}Tr_A(\Pi_{\pm}^A\otimes I^B\rho\Pi_{\pm}^A\otimes I^B),\nonumber\\
&=\left(\begin{array}{cc}
\rho_{11}+\rho_{33} & \pm(\rho_{14}+\rho_{23}) \\
\pm(\rho_{14}+\rho_{23}) & \rho_{22}+\rho_{44}
\end{array}\right),\nonumber\\
&=\frac{1}{2}
\left(\begin{array}{cc}
1+y & \pm t_1 \\
\pm t_1 & 1-y
\end{array}\right),
\end{align}
with $p_k=Tr(\Pi_{\pm}^A\otimes I^B\rho\Pi_{\pm}^A\otimes I^B)=\frac{1}{2}$, $k=\pm$ and
$y$, $t_1$ defined in Eqs. (\ref{parameter}).
In addition, $\rho_{\pm}^B$ have exactly the same spectrum
\begin{eqnarray}
\lambda(\rho_k)=\frac{1}{2}\left(1\pm\sqrt{y^2+t_1^2}\right),
\end{eqnarray}
Therefore
\begin{align}
\sum_kp_kS(\rho^B_k)&=S(\rho_+^B)=S(\rho_-^B),\nonumber\\
&=f(y^2+t_1^2).
\end{align}
where $f(z):=-\frac{1+\sqrt{z}}{2}\log_2\frac{1+\sqrt{z}}{2}-\frac{1-\sqrt{z}}{2}\log_2\frac{1-\sqrt{z}}{2}$.
If we choose $\{\Pi_{k}^A=\frac{1}{2}(I\pm\sigma_y)\}$ as the local measurement on A, an analogous expression
can be achieved
\begin{eqnarray}
\sum_kp_kS(\rho^B_k)=f(y^2+t_2^2),
\end{eqnarray}
According to Ref. \cite{Chen2011,Yu2011}, as long as
\begin{eqnarray}
|\sqrt{\rho_{11}\rho_{44}}-\sqrt{\rho_{22}\rho_{33}}|\leq|\rho_{14}|+|\rho_{23}|,
\end{eqnarray}
holds, the optimal observable is $\sigma_x$ if $t_1\geq t_2$ and $\sigma_y$ otherwise. Consequently,
\begin{eqnarray}
\mathcal{D}_A(\rho)=S(\rho^A)-S(\rho^{AB})+f(y^2+\max\{t_{1,2}^2\}),
\end{eqnarray}
Note that $S(\rho^A)$ and $S(\rho^{AB})$ can also be represented by parameters defined in
Eqs. (\ref{parameter}) (see Eq. (8) in Ref. \cite{Yu2011}).

%%%%%%%%%%%%%%%%%%%%%%%%%%%%%%%%%%%%%%%%%%%%%%%%%%%%%%%%%%%%%%%%%%%%%%%%%%%%%%%%%%%%%%%%%%%%%


\begin{thebibliography}{99}
\bibitem{EPR} A. Einstein, B. Podolsky, and N. Rosen, Phys. Rev. \textbf{47}, 777 (1935).
\bibitem{Nielsen} M. A. Nielsen and I. L. Chuang, Quantum Computation and Quantum Communication
(Cambridge University Press, Cambridge, 2000).
\bibitem{Ollivier2001} H. Ollivier and W. H. Zurek, Phys. Rev. Lett. \textbf{88}, 017901 (2001).
\bibitem{Henderson2001} L. Henderson and V. Vedral, J. Phys. A \textbf{34}, 6899 (2001).
\bibitem{Dakic2010} B. Daki\'{c}, V. Vedral, and C. Brukner, Phys. Rev. Lett.
\textbf{105}, 190502 (2010);
\bibitem{Luo2010a} S. Luo and S. Fu, Phys. Rev. A \textbf{82}, 034302 (2010).
\bibitem{Luo2008a} S. Luo, Phys. Rev. A \textbf{77}, 022301 (2008).
\bibitem{Luo2011} S. Luo and S. Fu, Phys. Rev. Lett. \textbf{106}, 120401 (2011).
\bibitem{Wu2009} S. Wu, U. V. Poulsen, and K. M{\o}lmer, Phys. Rev. A \textbf{80}, 032319 (2009).
\bibitem{Modi2011} K. Modi, A. Brodutch, H. Cable, T. Paterek, V. Vedral, arXiv:1112.6238.
\bibitem{Sachdev} S. Sachdev, Quantum Phase Transitions (Cambridge University
Press, Cambridge, 2000).
\bibitem{Osterloh2002} A. Osterloh, L. Amico, G. Falci, and R. Fazio, Nature (London) \textbf{416}, 608 (2002).
\bibitem{Osborne2002} T. J. Osborne and M. A. Nielsen, Phys. Rev. A \textbf{66}, 032110 (2002).
\bibitem{Vidal2003} G. Vidal, J. I. Latorre, E. Rico, and A. Kitaev, Phys. Rev. Lett. \textbf{90}, 227902 (2003).
\bibitem{Wu2004} L.-A. Wu, M. S. Sarandy, and D. A. Lidar, Phys. Rev. Lett. \textbf{93}, 250404 (2004).
\bibitem{Amico2008} L. Amico, R. Fazio, A. Osterloh, and V. Vedral, Rev. Mod. Phys. \textbf{80}, 517 (2008).
\bibitem{Dillenschneider2008} R. Dillenschneider, Phys. Rev. B \textbf{78}, 224413 (2008).
\bibitem{Sarandy2009} M. S. Sarandy, Phys. Rev. A \textbf{80}, 022108 (2009).
\bibitem{Werlang2010} T. Werlang, C. Trippe, G. A. P. Ribeiro, and G. Rigolin, Phys. Rev. Lett. \textbf{105}, 095702 (2010).
\bibitem{Wilson1975} K. G. Wilson, Rev. Mod. Phys. \textbf{47}, 773 (1975).
\bibitem{Pfeuty1982} P. Pfeuty, R. Jullian, K. L. Penson, in Real-Space Renormalization, edited by T. W. Burkhardt and J. M. J. van Leeuwen
(Springer, Berlin, 1982), Chap. 5.
\bibitem{Kargarian2007} M. Kargarian, R. Jafari, and A. Langari, Phys. Rev. A \textbf{76}, 060304 (2007).
\bibitem{Kargarian2008} M. Kargarian, R. Jafari, and A. Langari, Phys. Rev. A \textbf{77}, 032346 (2008).
\bibitem{Jafari2008} R. Jafari, M. Kargarian, A. Langari, and M. Siahatgar, Phys. Rev. B \textbf{78}, 214414 (2008).
\bibitem{Kargarian2009} M. Kargarian, R. Jafari, and A. Langari, Phys. Rev. A \textbf{79}, 042319 (2009).
\bibitem{Ma2011a} F. W. Ma, S. X. Liu, and X. M. Kong, Phys. Rev. A \textbf{83}, 062309 (2011).
\bibitem{Ma2011b} F. W. Ma, S. X. Liu, and X. M. Kong, Phys. Rev. A \textbf{84}, 042302 (2011).
\bibitem{Bell1964} J. S. Bell, Physics (Long Island City, N.Y.) \textbf{1}, 195 (1964).
\bibitem{CHSH1969} J. F. Clauser, M. A. Horne, A. Shimony, and R. A. Holt, Phys. Rev. Lett. \textbf{23}, 880 (1969).
\bibitem{Martin1996a} M. A. Martin-Delgado and G. Sierra, Int. J. Mod. Phys. A \textbf{11}, 3145 (1996).
\bibitem{Langari1998} A. Langari, Phys. Rev. B \textbf{58}, 14467 (1998); \textbf{69}, 100402(R) (2004);
\bibitem{Jafari2006} R. Jafari and A. Langari, Phys. Rev. B \textbf{76}, 014412 (2007); Physica A \textbf{364}, 213 (2006).
\bibitem{Gu2003} S.-J. Gu, H.-Q. Lin, and Y.-Q. Li, Phys. Rev. A \textbf{68}, 042330 (2003).
\bibitem{Gu2005} S.-J. Gu, G.-S. Tian and H.-Q. Lin, Phys. Rev. A \textbf{71}, 052322 (2005).
\bibitem{Martin1996b} M. A. Martin-Delgado and G. Sierra, Phys. Rev. Lett. \textbf{76}, 1146 (1996).
\bibitem{Wootters1998} W. K. Wootters, Phys. Rev. Lett. \textbf{80}, 2245 (1998)
\bibitem{Maziero2010} J. Maziero, T. Werlang, F. F. Fanchini, L. C. Celeri, and R. M. Serra, Phys. Rev. A \textbf{81}, 022116 (2010).
\bibitem{Fanchini2010} F. F. Fanchini, T. Werlang, C. A. Brasil, L. G. E. Arruda, and A. O. Caldeira, Phys. Rev. A \textbf{81}, 052107 (2010).
\bibitem{Ali2010} M. Ali, A. R. P. Rau, and G. Alber, Phys. Rev. A \textbf{81}, 042105 (2010).
\bibitem{Chen2011} Q. Chen, C. Zhang, S. Yu, X. X. Yi, and C. H. Oh, Phys. Rev. A \textbf{84}, 042313 (2011).
\bibitem{Yu2011} S. Yu, C. Zhang, Q. Chen, and C. H. Oh, arXiv:1102.1301.
\bibitem{Lu2011} X.-M. Lu, J. Ma, Z. Xi, and X. Wang, Phys. Rev. A \textbf{83}, 012327 (2011).
\bibitem{Gisin1991} N. Gisin, Phys. Lett. A \textbf{154}, 201 (1991).
\bibitem{Werner1989} R. F. Werner, Phys. Rev. A \textbf{40}, 4277 (1989).
\bibitem{Horodecki1995} R. Horodecki, P. Horodecki and M. Horodecki, Phys. Lett. A \textbf{200}, 340 (1995);
R. Horodecki, Phys. Lett. A \textbf{210}, 223 (1996).
\bibitem{Batle2010} J. Batle and M. Casas, Phys. Rev. A \textbf{82}, 062101 (2010).
\bibitem{Justino2012} L. Justino and Thiago R. de Oliveira, Phys. Rev. A \textbf{85}, 052128 (2012).

\end{thebibliography}
\end{document}